\documentclass[reprint,superscriptaddress,nofootinbib,amsmath,amssymb,aps,prd]{revtex4-1}

\usepackage{graphicx}
\usepackage{dcolumn}
\usepackage{bm}
\usepackage{booktabs}  
\usepackage{multirow}
\usepackage{longtable}
\usepackage{tabu}
\usepackage{appendix}

\usepackage{hyperref}
\hypersetup
{
colorlinks=true,
linkcolor=blue,
filecolor=blue,
urlcolor=blue,
citecolor=blue,
}

\begin{document}

\title{Quasi-normal modes of slowly-rotating Johannsen black holes}

\author{Yuhao~Guo}
\email{yhguo21@m.fudan.edu.cn}
\affiliation{Center for Astronomy and Astrophysics, Center for Field Theory and Particle Physics, and Department of Physics,\\
Fudan University, Shanghai 200438, China}

\author{Swarnim~Shashank}
\email{swarnim@fudan.edu.cn}
\affiliation{Center for Astronomy and Astrophysics, Center for Field Theory and Particle Physics, and Department of Physics,\\
Fudan University, Shanghai 200438, China}

\author{Cosimo~Bambi}
\email[Corresponding author: ]{bambi@fudan.edu.cn}
\affiliation{Center for Astronomy and Astrophysics, Center for Field Theory and Particle Physics, and Department of Physics,\\
Fudan University, Shanghai 200438, China}
\affiliation{School of Natural Sciences and Humanities, New Uzbekistan University, Tashkent 100007, Uzbekistan}

\date{\today}

\begin{abstract}
The detection of gravitational waves with ground-based laser interferometers has opened a new window to test and constrain General Relativity (GR) in the strong, dynamical, and non-linear regime. In this paper, we follow an agnostic approach and we study the quasi-normal modes of gravitational perturbations of Johannsen black holes under the assumptions of the validity of the Einstein Equations and of low values of the black hole spin parameter and deformation parameters. We find that the deformation parameter $\alpha_{13}$ has a stronger impact on the quasi-normal modes than the other leading order deformation parameters ($\alpha_{22}$, $\alpha_{52}$, and $\epsilon_{3}$). We derive a fitting formula for the fundamental modes with $l=2$ and $l=3$ for the deformation parameter $\alpha_{13}$ valid in the slow rotation approximation ($a_* < 0.4$). Finally, we constrain $\alpha_{13}$ from the event GW170104; within our analysis, we find that the data of GW170104 are consistent with the predictions of GR.
\end{abstract}

\maketitle


\section{Introduction}

The theory of General Relativity (GR) is one of the pillars of modern physics and all the available observations are consistent with its predictions\footnote{While there are problems, like those of dark matter and dark energy, none of these issues unambiguously requires modifications of GR and new physics may instead come from the matter sector.}. For decades, the theory has been mainly tested in the so-called {\it weak field regime} with experiments in the Solar System and observations of binary pulsars~\cite{Will:2014kxa}. On the contrary, the {\it strong field regime} was almost completely unexplored until 7-8~years ago. Thanks to a new generation of observational facilities, the situation has radically changed in the past years. Today we can test GR in the strong field regime with observations of black holes (BHs) from gravitational wave laser interferometers~\cite{LIGOScientific:2016lio,Yunes_2016,LIGOScientific:2020tif,Cardenas-Avendano:2019zxd,Shashank:2021giy}, X-ray missions~\cite{Cao:2017kdq,Tripathi:2018lhx,Tripathi:2020dni,Tripathi:2020yts}, and mm very long baseline interferometry facilities~\cite{Bambi:2019tjh,EventHorizonTelescope:2020qrl,EventHorizonTelescope:2022xqj,Vagnozzi:2022moj}. Current tests can be significantly improved in the future with the third generation of ground-based laser interferometers, space-based laser interferometers, upcoming X-ray observatories with larger effective areas and higher energy resolutions, and space very long baseline interferometry.

To test a theory, we can normally follow two different strategies, which are usually referred to as top-down (or theory-specific) approach and bottom-up (or agnostic) approach, respectively. In the case of tests of GR, both approaches have been employed, but the agnostic strategy is somewhat more popular than the theory-specific one. In the theory-specific approach, we want to test GR against another theory of gravity: we analyze some observations within GR and within the other theory of gravity and we check which of the two theories can explain the data better. To choose this strategy, we have to be able to derive the predictions of both GR and the other theory of gravity. With the agnostic approach, we develop a framework in which we can quantify possible deviations from the predictions of GR, usually by introducing some extra parameters. From the analysis of observations within this framework, we can measure the values of these extra parameters and check {\it a posteriori} if the latter are consistent with the values expected in GR.

BHs are ideal laboratories for testing GR in the strong field regime, as they are the sources of the strongest gravitational fields that we can find today in the Universe. In 4-dimensional GR and in the absence of exotic matter fields, uncharged BHs are described by the Kerr solution~\cite{Chrusciel:2012jk}. The spacetime geometry around astrophysical BHs is thought to be approximated well by the Kerr solution~\cite{Bambi:2017khi}. Testing the Kerr metric around astrophysical BHs is currently one of the most popular tests of GR in the strong field regime~\cite{Bambi:2015kza,Yagi:2016jml}. With the currently available data, X-ray tests of stellar-mass BHs in X-ray binary systems normally provide the most stringent constraints on possible deviations from the Kerr geometry~\cite{Bambi:2021chr,Bambi:2022dtw}. Gravitational wave (GW) tests of the Kerr metric are somewhat weaker than X-ray tests, but they promise to improve much faster in the next years.

Current GW observatories can detect the GW signal of the coalescence of two stellar-mass BHs. The coalescence process can be separated into three stages: the inspiral, the merger, and the ring-down. In the inspiral phase, the two objects rotates around a common center of mass and the system loses energy and angular momentum by emitting GWs. The inspiral signal is relatively well detected by current GW observatories and it is the most widely used to test GR; see, for instance, Refs.~\cite{Yunes_2016,Shashank:2021giy,Berti_2018,Shashank:2023erf,Shashank:2024zdr,Benavides-Gallego:2024hck,Das:2024mjq,Riaz:2022rlx}. In the merger, the two objects merge together to form a heavier BH: the merger signal could be potentially very useful to test GR, but it is too fast and short to be detected by current GW observatories and there are still large uncertainties in the theoretical predictions, even in GR. During the ring-down, the newly born BH emits GWs to settle down to an equilibrium configuration. In the ring-down phase, we have the emission of characteristic GW signals, which are called the quasi-normal modes (QNMs). The QNM signal is not measured very well with the sensitivity of current GW observatories, but it is very interesting for future tests of GR because the properties of the QNMs are independent of the initial conditions and only determined by the characteristic of the system.

There is a rich literature on the QNMs of BHs and neutron stars in GR and in other theories of gravity~\cite{Berti_2009,PANI_2013,Kokkotas_1999,Cardoso_2009,Konoplya_2011,Berti_2015,Barack_2019,berti2019tests} and the detection of QNMs promises to be a powerful tool for testing GR with future observations~\cite{Maggio_2023,Maselli_2020}. QNMs of specific gravity theories have been calculated~\cite{Pierini_2021,Bl_zquez_Salcedo_2016,Wagle_2022} and the available GW data from the LIGO-Virgo-KAGRA (LVK) Collaboration can already constrain some models~\cite{Silva:2022srr}. There are also attempts to use QNMs for agnostic tests of GR; see, for instance, Refs.~\cite{Volkel:2020daa,Volkel:2019muj}.

The properties of the QNMs can be inferred from the study of the evolution of small perturbations on a background metric. If we consider the QNMs of the perturbations of a scalar field, we have to solve the Klein-Gordon Equations in the BH metric that we want to study~\cite{Carson:2020iik,Carson:2020ter}. In GR, the QNM spectrum of the perturbations of a scalar field is similar to that of GWs, but in general this is not guaranteed. If we want to study the QNMs of gravitational perturbations, we need to know both the BH metric and the field equations of the gravity theory, and therefore, strictly speaking, this can be done only for testing specific gravity theories. Assuming that the GW emission can be approximated well by the Einstein Equations, we can potentially study the QNM spectrum of GWs on an arbitrary BH spacetime. However, even with such an assumption it is definitively non-trivial if the spacetime metric is only axisymmetric rather than spherically symmetric. The Master Equations of the QNMs can be further simplified if we consider slowly-rotating BHs, we expand the metric around the non-rotating solution, and we ignore all terms higher than the first order in the BH spin parameter.

In this paper, we study the QNMs of the gravitational perturbations in the Johannsen metric~\cite{Johannsen_2013} assuming the validity of the Einstein Equations and considering the slow rotation approximation. The Johannsen metric is a parametric BH spacetime specifically designed to test the Kerr hypothesis with astrophysical observations. It is characterized by four infinite sets of {\it deformation parameters}, which are introduced to quantify possible deviations from the Kerr background. If all deformation parameters vanish, we exactly recover the Kerr solution. The Johannsen metric is not a solution of any specific gravity theory, but it is supposed to approximate well a large number of BH solutions for a proper choice of the values of its deformation parameters. The use of the Johannsen metric assuming the validity of the Einstein Equations can make sense if we see our test as a null experiment: we expect that GR is correct and any small deviations from GR may be interpreted as a non-vanishing deformation parameter. In the case of a possible measurement of a non-vanishing deformation parameter, the method is clearly unsuitable to figure out the exact origin of the deviation from the GR predictions (which can be in the spacetime metric as well as in the generation of GWs), but it could still be an important first step before a more detailed (and model-dependent) analysis.

The Johannsen metric has been extensively used in X-ray tests of GR~\cite{Cao:2017kdq,Tripathi:2018lhx,Tripathi:2020dni,Tripathi:2020yts} and its leading deformation parameters have been already constrained from the inspiral signal of the GW events observed by the LVK Collaboration~\cite{Cardenas-Avendano:2019zxd,Shashank:2021giy,Das:2024mjq}. In this paper, we limit our study to the leading order deformation parameters and we derive a fitting formula. Thanks to the slow rotation approximation, axial and polar perturbations are separable. We use the direct integration method to calculate the QNMs~\cite{PANI_2013}. To test our method, we use the fitting formula to constrain one of the deformation parameters with the ring-down data of the LVK event GW170104.

The manuscript is organized as follows. In Section~\ref{s-jm}, we review the Johannsen metric. In Section~\ref{s-bhpt}, we derive the deformed Regge-Wheeler and Zerlini Equations for slowly-rotating Johannsen BHs. In Section~\ref{s-qnm0} and Section~\ref{s-qnmjm}, we calculate the QNMs in the Johannsen metric and we derive a fitting formula. In Section~\ref{s-datafit}, we obtain constraints on the Johannsen deformation parameter $\alpha_{13}$ from the ring-down data of GW170104. We discuss our results in Section~\ref{s-c}. In this manuscript we use natural units in which $c = G_{\rm N} = 1$.


\section{Johannsen metric}\label{s-jm}

In this section, we briefly review the Johannsen metric~\cite{Johannsen_2013}, which is a parametric BH spacetime specifically designed to test the Kerr hypothesis with astrophysical observations within an agnostic approach. In Boyer-Lindquist coordinates, the non-vanishing metric coefficients of the Johannsen metric are  
\begin{eqnarray}
     g_{tt}&=&-\frac{\Sigma[\Delta-a^{2}A_{2}^{2}(r)\sin^{2}\theta]}{[(r^{2}+a^{2})A_{1}(r)-a^{2}A_{2}(r)sin^{2}\theta]^{2}} \, , \nonumber \\
     g_{t\phi}&=&-\frac{a[(r^{2}+a^{2})A_{1}(r)A_{2}(r)-\Delta]\Sigma sin^{2}\theta}{[(r^{2}+a^{2})A_{1}(r)-a^{2}A_{2}(r)\sin^{2}\theta]^{2}} \, ,\nonumber \\
     g_{rr}&=&\frac{\Sigma}{\Delta A_{5}(r)}\, , \nonumber \\
     g_{\theta\theta}&=&\Sigma \, , \nonumber\\
     g_{\phi\phi}&=&\frac{\Sigma \sin^{2}\theta[(r^{2}+a^{2})^{2}A_{1}^{2}(r)-a^{2}\Delta \sin^{2}\theta]}{[(r^{2}+a^{2})A_{1}(r)-a^{2}A_{2}(r)\sin^{2}\theta]^{2}} \, ,
\end{eqnarray}
where $\Sigma = r^2 + a^2 \cos^2\theta + f(r)$, $\Delta = r^2 - 2 M r + a^2$, and the functions $A_{1}(r)$, $A_{2}(r)$, $A_{5}(r)$, and $f(r)$ are given by
\begin{eqnarray}
A_{1}(r)&=&1+\sum_{n=3}^{\infty} {\alpha_{1n}\left(\frac{M}{r}\right)^{n}} \, , \nonumber\\
A_{2}(r)&=&1+\sum_{n=2}^{\infty} {\alpha_{2n}\left(\frac{M}{r}\right)^{n}} \, , \nonumber\\
A_{5}(r)&=&1+\sum_{n=2}^{\infty} {\alpha_{5n}\left(\frac{M}{r}\right)^{n}} \, , \nonumber\\
f(r)&=&\sum_{n=3}^{\infty} {\epsilon_{n}\frac{M^{n}}{r^{n-2}}} \, .
\end{eqnarray}
With this version of the metric, the spacetime is asymptotically flat, we recover the correct Newtonian limit, and there are no constraints from Solar System experiments~\cite{Johannsen_2013}.

In what follows, we restrict our study to the leading order deformation parameters in $A_{1}$, $A_{2}$, $A_{5}$, $f(r)$, namely $\alpha_{13}$, $\alpha_{22}$, $\alpha_{52}$, and $\epsilon_{3}$, and all higher order terms are assumed to vanish. Moreover, we consider slowly-rotating BHs and we keep only the terms linear in the BH spin $a$. In this context, the non-vanishing metric coefficients of the Johannsen metric become
\begin{eqnarray}\label{eq-jm-1}
g_{tt}&=&-\frac{\Delta  \Sigma }{(A_{1} r^{2})^{2}} \, , \nonumber \\
g_{t\phi}&=&-\frac{a \Sigma \sin^{2}\theta (A_{1}A_{2} r^2-\Delta) }{(A_{1} r^{2})^{2}} \, , \nonumber \\
g_{rr}&=&\frac{\Sigma }{A_{5} \Delta } \, , \nonumber \\
g_{\theta\theta}&=&\Sigma \, , \nonumber \\
g_{\phi\phi}&=&\Sigma  \sin^{2}\theta \, , 
\end{eqnarray}
where
\begin{eqnarray}\label{eq-jm-2}
A_{1}(r)&=&1+\alpha_{13}\left(\frac{M}{r}\right)^{3} \, , \nonumber\\
A_{2}(r)&=&1+\alpha_{22}\left(\frac{M}{r}\right)^{2} \, , \nonumber\\
A_{5}(r)&=&1+\alpha_{52}\left(\frac{M}{r}\right)^{2} \, , \nonumber\\
\Sigma&=&r^{2}+\epsilon_{3} \frac{M^{3}}{r} \, .
\end{eqnarray}
We note that the slow rotation approximation is adopted in order to separate axial and polar perturbations.


\section{Master equations}\label{s-bhpt}

We follow the standard approach in BH perturbation theory; see, for instance, Ref.~\cite{Wagle_2022}. We consider linear perturbations on the background metric
\begin{equation}\label{eq-mbp}
g_{\mu\nu} = g_{\mu\nu}^{0} + h_{\mu\nu} \,, 
\end{equation} 
where, in our case, $g_{\mu\nu}^{0}$ is the Johannsen metric given in Eqs.~(\ref{eq-jm-1}) and (\ref{eq-jm-2}), and $|h_{\mu\nu}| \ll 1$. The angular dependence of the perturbations $h_{\mu\nu}$ can be decomposed into scalar, vector, and rank-2 tensor spherical harmonics. Vector and tensor spherical harmonics are grouped into two classes, according to their behavior under a parity transformation ($\theta \rightarrow \pi -\theta$ and $\phi \rightarrow \phi +\pi$), and we can write $h_{\mu\nu} = h_{\mu\nu}^{\rm polar} + h_{\mu\nu}^{\rm axial}$. Polar/even perturbations get the factor $(-1)^{l}$ under a parity transformation. Axial/odd perturbations get the factor $(-1)^{l+1}$.

\begin{widetext}
We choose the Regge-Wheeler gauge to simplify the calculations~\cite{Regge:1957td}. Polar and axial perturbations can now be written as 
\begin{eqnarray}
    h_{\mu\nu}^{\rm polar}
    &=& \sum_{l=0}^{\infty} \sum_{m=-l}^{l}
    \left(
    \begin{array}{cccc}
     H_{0}^{lm}&H_{1}^{lm}&0&0\\
     H_{1}^{lm}&H_{2}^{lm}&0&0\\
     0&0&r^{2} K^{lm}&0\\
     0&0&0&r^{2} \sin^{2}\theta K^{lm}
    \end{array}
    \right) Y^{lm} \, , \\ 
        h_{\mu\nu}^{\rm axial}
    &=& \sum_{l=0}^{\infty} \sum_{m=-l}^{l}
    \left(
    \begin{array}{cccc}
     0&0& -\frac{h_{0}^{lm}}{\sin\theta} \partial_{\phi} & h_{0}^{lm} \sin\theta \partial_{\theta} \\
     0&0& -\frac{h_{1}^{lm}}{\sin\theta} \partial_{\phi} & h_{1}^{lm} \sin\theta \partial_{\theta} \\
     -\frac{h_{0}^{lm}}{\sin\theta} \partial_{\phi} & -\frac{h_{1}^{lm}}{\sin\theta} \partial_{\phi} &0&0\\
     h_{0}^{lm} \sin\theta \partial_{\theta} & h_{1}^{lm} \sin\theta \partial_{\theta} &0&0
    \end{array}
    \right) Y^{lm} \, , \nonumber \\
\end{eqnarray}
\end{widetext}
where $H^{lm}_i = H^{lm}_i (t,r)$, $K^{lm} = K^{lm} (t,r)$, and $h^{lm}_i = h^{lm}_i (t,r)$ are functions only of the coordinates $t$ and $r$, while $Y^{lm} = Y^{lm} (\theta,\phi)$ are the scalar spherical harmonics and depend only on the coordinates $\theta$ and $\phi$.

We assume the Einstein Equations. We plug the metric in~(\ref{eq-mbp}) into the Einstein Equations and we ignore all terms of second or higher order in $h_{\mu\nu}$, all terms of second or higher order in the BH spin $a$, all terms of second or higher order in the deformation parameters ($\alpha_{13}$, $\alpha_{22}$, $\alpha_{52}$, and $\epsilon_3$). We assume the following form for the functions $H^{lm}_i$, $K^{lm}$, and $h^{lm}_i$ 
\begin{eqnarray}
H^{lm}_i (t,r) &=& e^{-i \omega t} \tilde{H}^{lm}_i (r) \, , \nonumber\\ 
K^{lm} (t,r) &=& e^{-i \omega t} \tilde{K}^{lm} (r) \, , \nonumber\\ 
h^{lm}_i (t,r) &=& e^{-i \omega t} \tilde{h}^{lm}_i (r) \, .
\end{eqnarray}

The Johannsen metric is not a vacuum solution of the Einstein Equations. If we plug the Johannsen metric into the left hand side of the Einstein Equations, we get a non-vanishing effective stress-energy tensor on the right hand side. If we consider gravitational perturbations $h_{\mu\nu}$, we may consider even perturbations in the effective stress-energy tensor, which can be decomposed to spherical harmonics too. Within our agnostic approach, we assume that there are no perturbations on the right hand side.

Within our approximation, we can separate the angular dependence of the perturbations from the radial one and, in turn, we can separate the radial dependence of polar and axial perturbations. Our Master Equations of our gravitational perturbations turn out to have the form of the Regge-Wheeler and Zerilli Equations
\begin{eqnarray}\label{eq-Master}
\mathcal{D} \Psi + V \Psi=0
\end{eqnarray}
where $\Psi$ is one of the functions $\tilde{H}^{lm}_i$, $\tilde{K}^{lm}$, and $\tilde{h}^{lm}_i$, $\mathcal{D}$ is a second order radial differential operator, and $V$ is an effective potential. In tortoise coordinates, the Master Equations reduce to the following form 
\begin{eqnarray}\label{eq-tortoise}
\frac{\partial^2 \Psi}{\partial r_*^2} + V_* \Psi=0 \, ,
\end{eqnarray}
where
\begin{eqnarray}
r_{*} = r + 2M \ln \left( \frac{r}{2M} - 1 \right) \, . 
\end{eqnarray}

In the case of axial perturbations ($\Psi = \tilde{h}^{lm}_1$, $\tilde{h}^{lm}_2$), the effective potential $V$ is
\begin{widetext}
\begin{eqnarray}
    V^{\rm axial}&=& \omega^2 + \frac{2 \left(l^2+l+3\right) M r - l (l + 1) r^2 -12 M^2}{r^4} \nonumber\\
    &&-\frac{4 m a M}{r^7 \omega} \frac{\left(l^2 r^4 \omega ^2+l r^4 \omega ^2+84 M^2-78 M r+18 r^2\right)}{l (l+1)} \nonumber \\
    &&+\frac{\alpha_{13} M^3}{r^8} \left\{ 
    \frac{3 (2 M-r) \left(-9 \left(l^2+l-8\right) r^2+(25 l (l+1)-428) M r+576 M^2\right)}{l^2+l-2}+2 r^5 \omega ^2\right\} \nonumber\\
    &&+ \frac{\alpha_{52} M^3}{r^7} \frac{84 M^2 -2 [4 l (l+1)+55] M r+[44-l (l+1) (2 l (l+1)-13)] r^2}{l^2+l-2} \nonumber \\
    &&+\frac{\alpha_{52} M^2}{r^4} \frac{l (l+1) \left[2 l (l+1)-2 r^2 \omega ^2-9\right]+4 r^2 \omega ^2-10}{2 \left(l^2+l-2\right)} \nonumber\\
    &&-\frac{3 \epsilon_{3} M^3 (2 M-r)}{r^8} \frac{\left(9 l^2+9 l-368\right) M r-3 \left(l^2+l-20\right) r^2+512 M^2}{l^2+l-2} \, .
\end{eqnarray}
\end{widetext}
In the Schwarzschild limit ($a = \alpha_{13} = \alpha_{22} = \alpha_{52} = \epsilon_3 = 0$), we have only the first two terms. For slowly-rotating Kerr BHs ($\alpha_{13} = \alpha_{22} = \alpha_{52} = \epsilon_3 = 0$), we have the first three terms. We note that the deformation parameter $\alpha_{22}$ does not appear in the effective potential within our approximations.

The effective potential for polar perturbations ($\Psi = \tilde{H}^{lm}_0$, $\tilde{H}^{lm}_1$, $\tilde{H}^{lm}_2$, , $\tilde{K}^{lm}$) has quite a long expression and is reported in Appendix~\ref{A}. As in the case of the effective potential for axial perturbations, the deformation parameter $\alpha_{22}$ is not present within our approximations. In the rest of the paper, we will thus consider only the deformation parameters $\alpha_{13}$, $\alpha_{52}$, and $\epsilon_3$.

In the cases of the Schwarzschild and Kerr spacetimes, polar and axial perturbations turn out to have the same spectrum. However, in general such a result is not guaranteed and the spectra of polar and axial perturbations may be different. In our case, since we are assuming that the deformation parameters are small, we can expect that any difference between the spectra of polar and axial perturbations is also small. In what follows, we study the axial perturbations assuming that, at first approximation, our results hold even for polar perturbations.


\section{Quasi-normal modes}\label{s-qnm0}

The late-time perturbations of the background metric are sinusoids known as the QNMs. These characteristic vibration modes of the spacetime have complex frequencies: the real part describes the oscillation of the metric perturbation while the imaginary part describes the decay of the amplitude of the oscillation. 

From the Master Equations derived in the previous section, we can calculate the frequencies of the QNMs with one of the available numerical methods discussed in the literature. In what follows, we use the direct integration method~\cite{PANI_2013}.

\subsection{Boundary conditions}

In our Master Equations, we have to impose two boundaries: one at spatial infinity ($r\rightarrow \infty$) and the other one at the BH event horizon ($r = r_{\rm H}$). Within our approximation of slow rotation and small deformation parameters, the radial coordinate of the BH event horizon reduces to that of a Schwarzschild BH, i.e. $r_{\rm H} = 2 M$. QNM solutions are purely in-going at the event horizon and purely out-going at spatial infinity:
\begin{eqnarray}
    \Psi\propto\left\{\begin{aligned}
        &e^{-i \omega_{H}r_{*}} \, , & r\rightarrow r_{\rm H} \, , \\
        &e^{i \omega_\infty r_{*}} \, , & r\rightarrow \infty \, ,   
    \end{aligned} \right.
\end{eqnarray}
where
\begin{eqnarray}\label{eq-omegah}
    \omega_{\rm H} &=& \sqrt{\omega  \left[\omega  (\alpha_{13}-\alpha_{52}+4)-\frac{2 m a}{M^2}\right]} \, , \nonumber\\
    \omega_\infty &=& \omega \, .
\end{eqnarray}

We note that, while we have derived the Master Equations ignoring terms of second and higher order in the perturbations, spin parameter, and deformation parameters, we will use the full expression in Eq.~(\ref{eq-omegah}) for $\omega_{\rm H}$. The exact value of $\omega_{\rm H}$ has indeed a strong impact on the calculations and we want to investigate deformation parameters of order of 0.1.

At the event horizon, $r = r_{\rm H}$ but $r_* \rightarrow -\infty$. At spatial infinity, both $r$ and $r_*$ diverge. In the numerical calculations of the QNMs, this can cause problems. It is thus convenient to find analytic asymptotic solutions to the perturbation equations around the BH event horizon and spatial infinity of the following form 
\begin{eqnarray}
    \Psi\propto \left\{ 
    \begin{aligned}
        &e^{-i\omega_{\rm H}r_{*}}\sum_{n=0}^{\infty} h_{n}^{i}(r- 2M)^{n} \, , & r\rightarrow r_{H} \, ,\\
        &e^{i\omega r_{*}} \sum_{n=0}^{\infty}  g_{n}^{i} r^{-n} \, , & r\rightarrow \infty \, ,
    \end{aligned} \right.
\end{eqnarray}
where the index $i$ is to distinguish coefficients of the same order for different deformation parameters. The coefficient $h_{n}^{i}$ and $g_{n}^{i}$ can be determined, order by order, from our Master Equations for $r\rightarrow r_{H}$ and $r\rightarrow \infty$, respectively. For our calculations, we calculate first terms to a maximum calculating limit, which is the maximum calculable order of expansion coefficients in Mathematica. In our case, we set n=10 for the boundary condition at the horizon and n=14 for the boundary condition at infinity.

\subsection{Direct integration method}

With the direct integration method, we directly integrate the Master Equations from the event horizon (in our numerical calculations, from $r_1 = r_{\rm H} + 0.0001 \; M$) to a finite radius ($r_{\rm matching} = 20 \; M$) and from infinity (in our numerical calculations, from $r_2 = 60 \; M$) to the same finite radius.

The Wronskian matrix of these two solutions is
\begin{eqnarray}
    W
    &=&
    \left(
    \begin{array}{cc}
     \Psi_{\rm in} & \Psi_{\rm in}' \\
      \Psi_{\rm out} & \Psi_{\rm out}'
    \end{array}
    \right)  
\end{eqnarray}
where the subscript in (out) indicates the solution of the integration from the event horizon (from infinity) to the finite radius and $\Psi' = \partial\Psi/\partial r$. The two solutions are the same solution when 
\begin{eqnarray}
    {\rm det}(W)|_{\rm matching}=0 
\end{eqnarray}
which is the condition for $\omega$ to be a frequency of a QNM. We use the {\tt DNSolve} function in {\tt Mathematica} to find the QNM frequencies and we set the accuracy and the precision to 10~digits.


\section{Results}\label{s-qnmjm}

The obtained $\omega$ is comprised of a real part, describing the frequency of the oscillation of the metric perturbation, and an imaginary part, describing the decay of the amplitude. We label the QNM by increasing $|{\rm Im}(\omega)|$ as overtone number. Among all QNMs, the dominant one is normally expected to be the $\{l,m,n\}=\{2,2,1\}$ mode, $\omega_{221}$~\cite{Berti:2005ys,Berti:2007zu,Berti:2009kk}. 
We choose this mode for our study. In reality, the GW signal from a certain astrophysical event should be the sum of a few dominant modes. Without loss of generality, in our calculations we set $M=1$.

\subsection{Kerr spacetime and correction factor}\label{appro}

With the metric decomposition presented in the previous section, the gravitational perturbations can be decomposed into a class of partial differential equations with the form in Eq.~(\ref{eq-Master}). However, in general, these equations are not separable unless the slow rotation approximation is adopted. In our case of the Johannsen metric with deformation parameters $\alpha_{13}$, $\alpha_{52}$, and $\epsilon_3$, we also need to consider these three deformation parameters to first order in order to have separable equations.

\begin{figure*}[t]
\centering
\includegraphics[width= 0.98\textwidth]{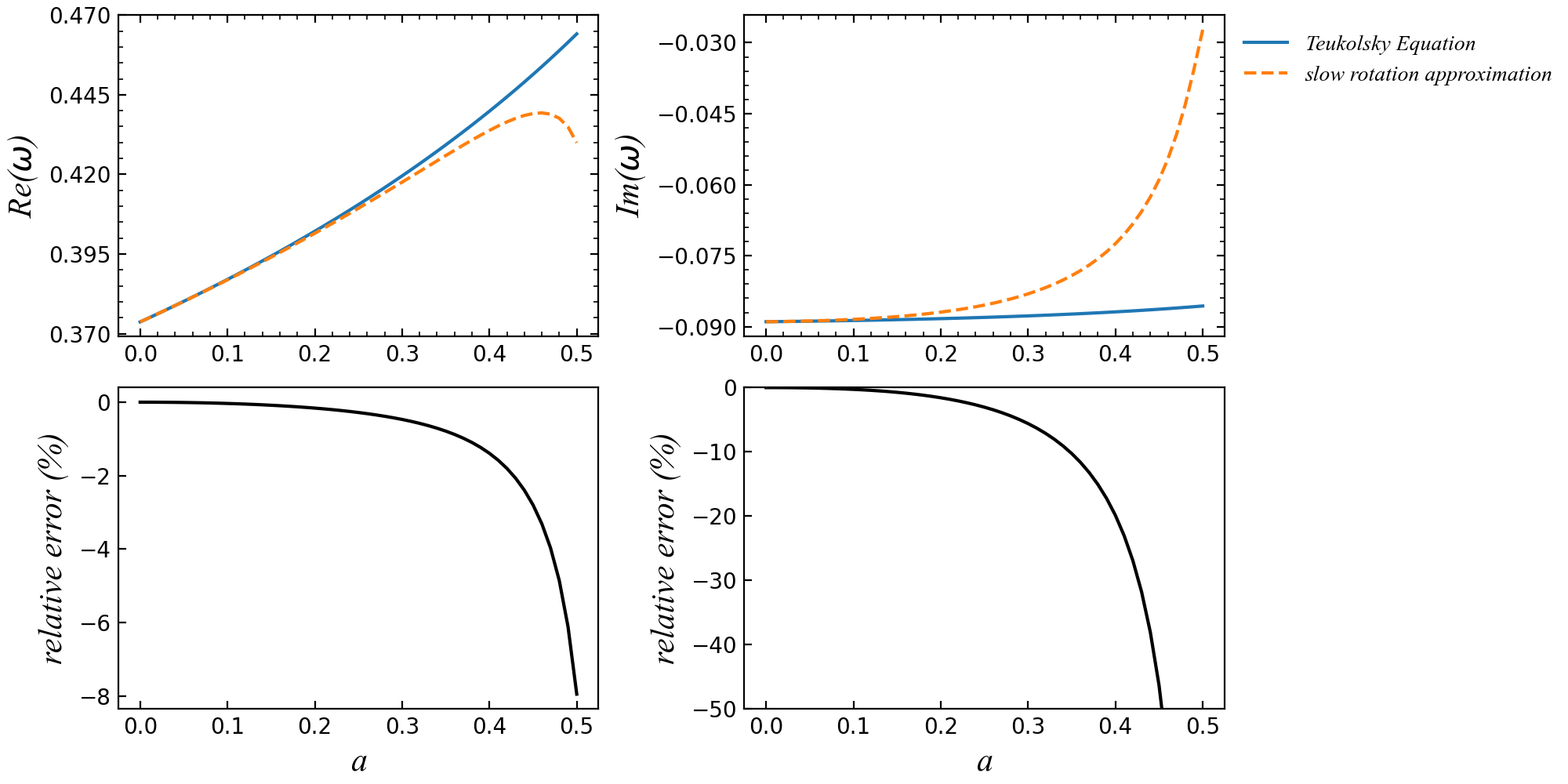}
\caption{Comparison between the real part (top left panel) and the imaginary part (top right panel) of the QNM $\{l,m,n\}=\{2,2,1\}$ in the Kerr spacetime as calculated through the Teukolsky Equations (blue-solid curves) and through our method with the slow rotation approximation (orange-dashed curves). The bottom panels show the relative error of our calculations. We use units in which $M=1$. See the text for more details. }\label{fig-sra}
\end{figure*}

Fig.~\ref{fig-sra} shows the real and imaginary parts of $\omega_{221}$ in the Kerr spacetime ($\alpha_{13}=\alpha_{52}=\epsilon_3=0$) as a function of the BH spin $a$ as calculated through our method with the slow rotation approximation and through the Teukolsky Equations (valid for any value of the BH spin parameter). The lower panels in Fig.~\ref{fig-sra} show the relative error of the real and imaginary parts of $\omega_{221}$ obtained with our method. For the real part, the error is below 1\% for $a < 0.3$ and it is about 1\% for $a=0.4$. For the imaginary part, the error is larger: it is below 10\% for $a < 0.3$ and it is about 20\% for $a=0.4$. We also note that our slow rotation approximation predicts that the real part of $\omega_{221}$ increases as the spin increases up to $a \approx 0.45$, and then decreases for higher values of $a$.

Astrophysical BHs are expected to rotate, just like the vast majority of astrophysical objects. In an attempt to improve the accuracy of our predictions, we introduce a correction factor $C$, which is calculated as the ratio between the QNM frequency in the Kerr spacetime calculated with the Teukolsky Equations, $\omega_{\rm Teu}$, and the QNM frequency in the Kerr spacetime calculated with our method, $\omega_{\rm Kerr}$: $C = \omega_{\rm Teu}/\omega_{\rm Kerr}$. The {\it corrected} QNM frequency $\omega_{\rm corr}$ is
\begin{eqnarray}
\omega_{\rm corr} = C \; \omega_{\rm Kerr} \, .
\end{eqnarray}
By definition of the correction factor $C$, $\omega_{\rm corr} = \omega_{\rm Teu}$ when we consider the Kerr spacetime. In the next sections, we will use the correction factor $C$ to improve the accuracy of our predictions of the QNM frequencies for slow-rotating non-Kerr black holes.

\begin{figure*}[t]
\centering
\includegraphics[width=0.98\textwidth]{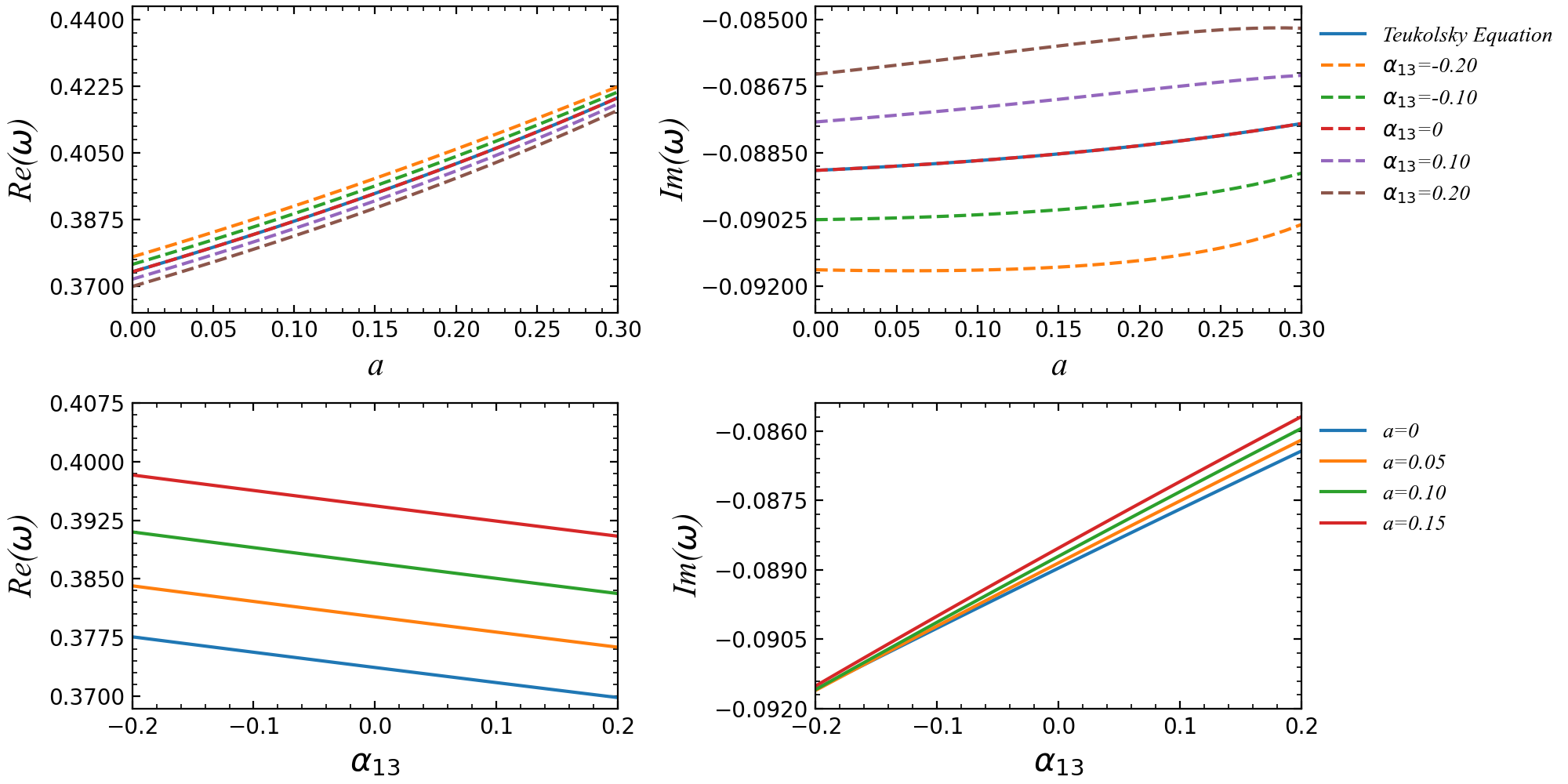}
\caption{QNM $\{l,m,n\}=\{2,2,1\}$ in the Johannsen spacetime with non-vanishing BH spin $a$ and deformation parameter $\alpha_{13}$. The top panels show the real and imaginary parts of the QNM frequency as a function of the BH spin $a$ for different values of the deformation parameter $\alpha_{13}$. The bottom panels show the real and imaginary parts of the QNM frequency as a function of the deformation parameter $\alpha_{13}$ for different values of the BH spin $a$. We use units in which $M=1$. See the text for more details.}\label{f-a13}
\end{figure*}

\subsection{Non-Kerr spacetime}

We can repeat our calculations of the QNM frequencies in the presence of a non-vanishing deformation parameter. If the QNM frequency calculated in the deformed Kerr spacetime with our method (slow rotation and small deviations from Kerr) is $\omega_{\rm Non-Kerr}$, the corrected QNM frequency is
\begin{eqnarray}
\omega_{\rm corr} = \omega_{\rm Non-Kerr} + \left(C - 1\right) \, \omega_{\rm Kerr} \, ,
\end{eqnarray}
where $\omega_{\rm Kerr}$ is the QNM frequency in the Kerr spacetime for the same value of the black hole spin parameter with our method. In this way, we exactly recover the result of the Teukolsky Equations in the limit of vanishing deformation parameters and we expect to have a more accurate QNM frequency in the limit of small deformation parameters.

Fig.~\ref{f-a13} shows $\omega_{\rm corr}$ for $a < 0.3$ and $-0.2 < \alpha_{13} < 0.2$ for the QNM $\{l,m,n\}=\{2,2,1\}$. A positive $\alpha_{13}$ decreases the value of the real part of the QNM frequency (left panels) and increases that of the imaginary part (right panels). We note that $\omega_{\rm Non-Kerr}$ depends even on the BH spin and therefore the $\omega_{\rm corr}$ curves in Fig.~\ref{f-a13} are not rigid vertical translations of $\omega_{\rm Teu}$ on the plane spin parameter vs $\alpha_{13}$.

\begin{figure*}
\label{fig:a52}
\centering
\includegraphics[width=0.98\textwidth]{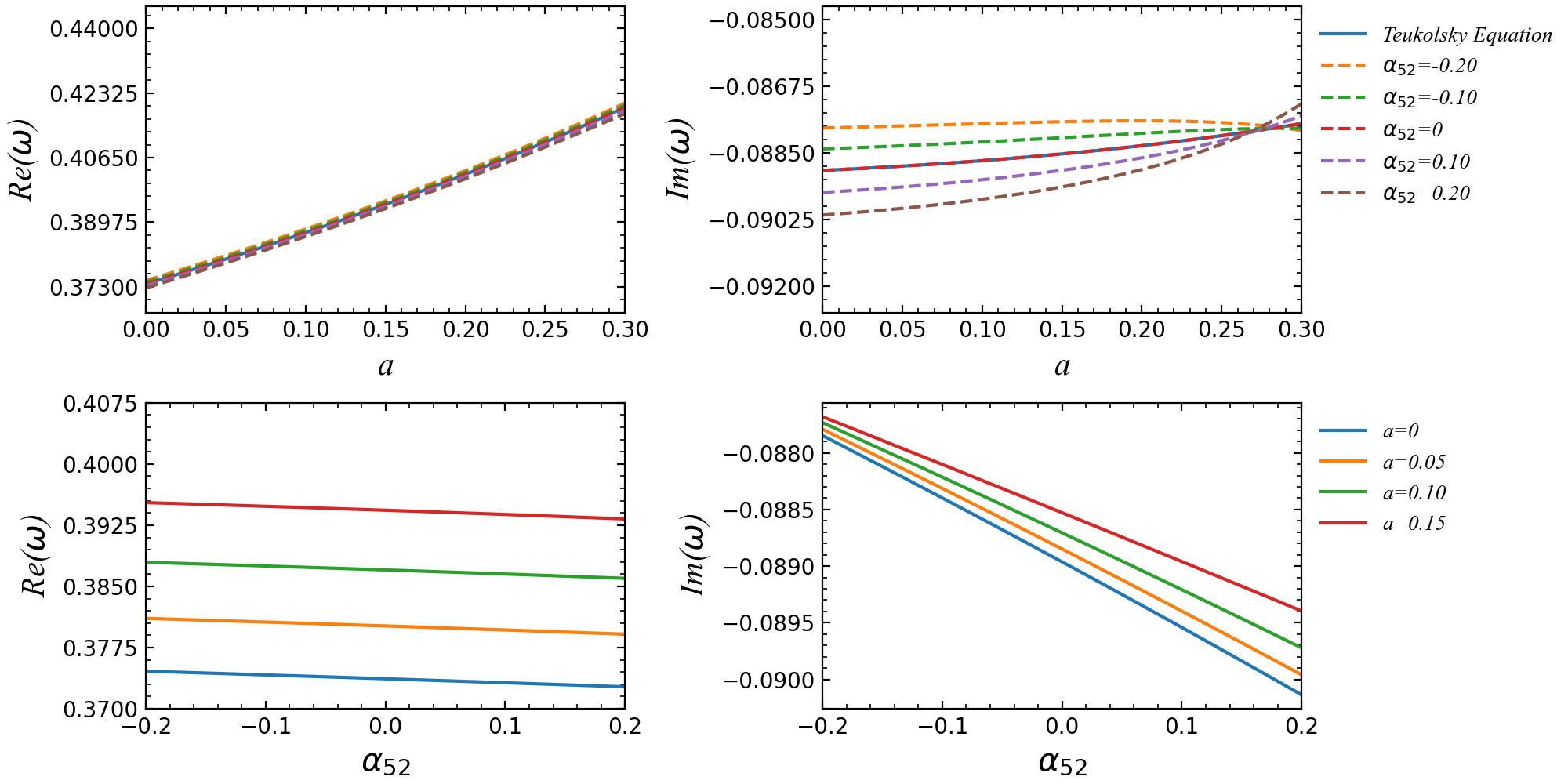}
\caption{As in Fig.~\ref{f-a13} for the deformation parameter $\alpha_{52}$. See the text for more details.}\label{f-a52}
\vspace{1.0cm}
\centering
\includegraphics[width=0.98\textwidth]{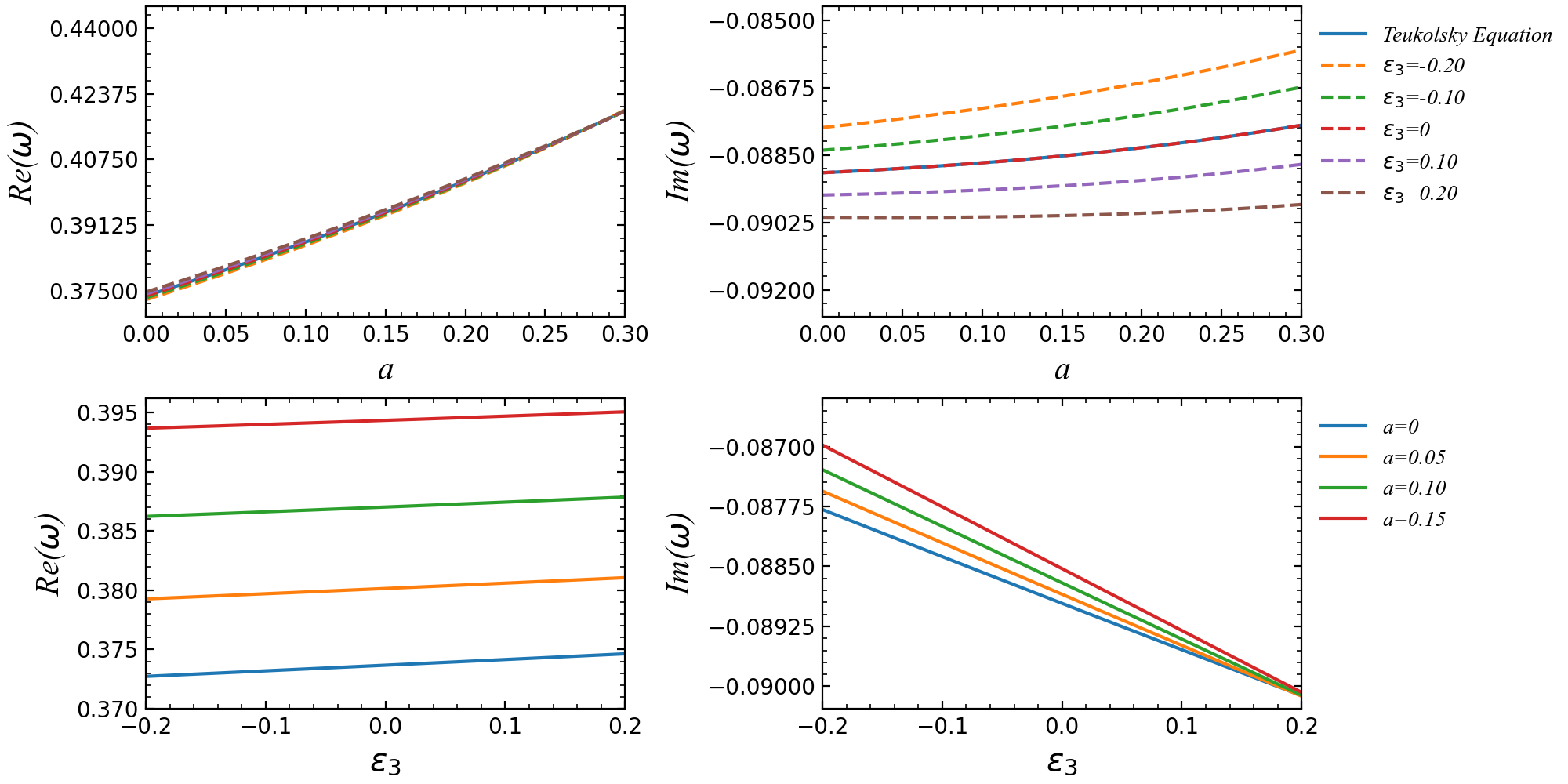}
\caption{As in Fig.~\ref{f-a13} for the deformation parameter $\epsilon_{3}$. See the text for more details.}\label{f-e3}
\end{figure*}

Fig.~\ref{f-a52} shows $\omega_{\rm corr}$ for $a < 0.3$ and $-0.2 < \alpha_{52} < 0.2$ for the QNM $\{l,m,n\}=\{2,2,1\}$. A positive $\alpha_{52}$ decreases both the values of the real and imaginary parts of the QNM frequency.

Fig.~\ref{f-e3} shows $\omega_{\rm corr}$ for $a < 0.3$ and $-0.2 < \epsilon_3 < 0.2$ for the QNM $\{l,m,n\}=\{2,2,1\}$. A positive $\epsilon_3$ increases the value of the real part of the QNM frequency (left panels) and decreases that of the imaginary part (right panels).


Among deviations come from three deformation parameters, $\alpha_{13}$ has the largest impact on the both parts QNM frequency under small spin approximation. Hence, we construct fitting formula with the dominant parameter $\alpha_{13}$ under slow spin range and ignore the impact of $\alpha_{52},\epsilon_{3}$.

\subsection{Fitting formula}
\label{sec:fitting_formula}

Now we construct a fitting formula for the real and imaginary parts of the QNM frequencies. We know the QNM frequencies in the Kerr spacetime calculated with the Teukolsky Equations and we look for a formula to take a non-vanishing $\alpha_{13}$ into account. Our fitting formula has the following form
\begin{eqnarray}
    Re(\omega)&=&Re(\omega_{\mathrm{Teu}})+P_{0}\alpha_{13}+P_{1}\alpha_{13}a+\mathcal{O}(a^{2},\alpha_{13}^{2}), \nonumber\\
    Im(\omega)&=&Im(\omega_{\mathrm{Teu}})+Q_{0}\alpha_{13}+Q_{1}\alpha_{13}a+\mathcal{O}(a^{2},\alpha_{13}^{2}). \nonumber\\
\end{eqnarray}

To derive the coefficients $P_0$, $P_1$, $Q_0$, and $Q_1$, we consider the ranges $a \in [0, 0.3]$ and $\alpha_{13} \in [-0.3, 0.3]$, where the slow spin approximation and the small deformation parameter approximation can accurately describe the QNM values. We use {\tt numpy.polyfit} to fit the coefficients~\cite{Harris:2020xlr} and we report our results for the dominant mode $\{l,m,n\}=\{2,2,1\}$ in Tab.~\ref{tab-fit}. The coefficients of the fitting formula for first overtone mode of $l=2$ and 3 are reported in Appendix~\ref{C}. The errors is not more than 0.2\% for $l=2$ and not more than 0.3\% for $l=3$.

\begin{table}[htbp]
\caption{Coefficients of the fitting formula for $\{l,m,n\}=\{2,2,1\}$.}\label{tab-fit}
\begin{ruledtabular}
\begin{tabular}{cccccc}
    & $P_{0}$ & $P_{1}$ & $Q_{0}$ & $Q_{1}$ &error [\%] \\ \hline
   Re($\omega_{221}$) & $-0.0197$ & 0.0025 & & & 0.18   \\
 Im($\omega_{221}$)  & & & 0.0130 & $-0.0095$ & 0.14  \\
\end{tabular}
\end{ruledtabular}
\end{table}


\section{Constraining $\alpha_{13}$ with GW data}\label{s-datafit}

The fitting formula allows us to constrain $\alpha_{13}$ using events whose spin lies in our low-spin band, $a \in [0, 0.4]$. For a given mode $\{l,m,n\}$, we can write \cite{pyRing,Carullo_2019,Isi_2019}:
\begin{eqnarray}
    f_{lmn}=f_{lmn}^{GR}(1+\delta f_{lmn}), \nonumber \\
    \tau_{lmn}=\tau_{lmn}^{GR}(1+\delta \tau_{lmn}) \, ,
\end{eqnarray}
where $f_{lmn}=Re(\omega_{lmn})$ and $\tau_{lmn}=1/Im(\omega_{lmn})$ are the frequency and damping time of the QNM $\{ n,l,m \}$ and $\delta f_{lmn}$ and $\delta \tau_{lmn}$ as their corresponding deviations from the GR prediction. A mapping can then be established between the fitting formula and the parametric GR deviation
\begin{eqnarray}
    \label{map}
    \alpha_{13} &=& \frac{f_{lmn}^{GR} \delta f_{lmn}}{P_{0}+P_{1} a}+\mathcal{O}(a^{2}), \nonumber \\
    &=& -\frac{ \delta \tau_{lmn}}{\tau_{lmn}^{GR}(1+\delta \tau_{lmn})(Q_{0}+Q_{1} a)}+\mathcal{O}(a^{2}).
\end{eqnarray}

\begin{figure}
    \centering
    \includegraphics[width=0.45\textwidth]{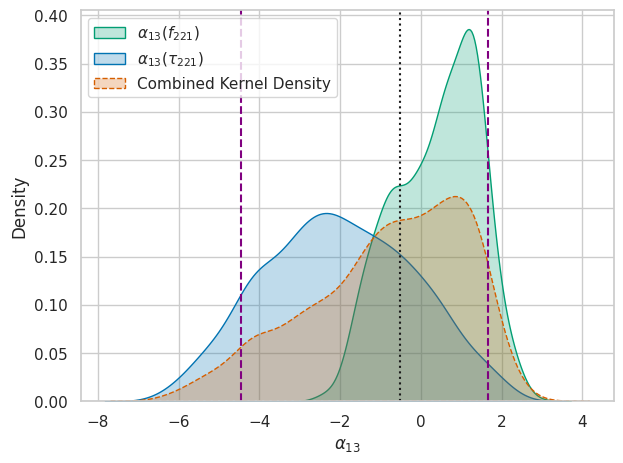}
    \caption{Constraints on $\alpha_{13}$ obtained from the $\{2,2,1\}$ mode of GW170104. We use a kernel density function to combine the two constraints obtained from $\delta f$ (green area) and $\delta \tau$ (blue area). The vertical black dotted line marks the median $-0.53$ and the two vertical violet dashed lines mark the 90\% confidence interval $[-4.46,1.66]$ of the combined constraint. }\label{f-fit}
\end{figure}

To fit the data, we choose the $\{2,2,1\}$ mode of event GW170104, as the spin estimate for this source is low (see Tab.~VIII in Ref.~\cite{LIGOScientific:2020tif}, ${\rm Kerr}_{220}$ in their convention). Using {\tt pyRing} \cite{pyRing,Carullo_2019,Isi_2019}, we obtain the posterior distributions of final mass, final spin, $\delta f_{221}$, and $\delta \tau_{221}$. Using Eq.~\ref{map}, the deformation parameter $\alpha_{13}$ is estimated, as shown in Fig.~\ref{f-fit}. The spin range we obtain is $a = 0.31^{+0.45}_{-0.28}$ at the 90\% confidence level (CL), which exceeds slightly the range allowed with our methods. Nevertheless, this is the best we have for now to estimate $\alpha_{13}$ from real data. The posteriors obtained for $\delta f_{221}$ is $-0.15^{+0.54}_{-0.32}$ and for $\delta \tau_{221}$ is $-0.30^{+0.51}_{-0.18}$, both at 90\% CL.

Given the range of values for the posteriors, we extend the fitting formula in Section~~\ref{sec:fitting_formula} for $\alpha_{13} \in [-1.0, 1.0]$ and obtain $P_0 = -0.0205$, $P_1 = 0.0166$, $Q_0 = 0.0136$, and $Q_1 = 0.0021$. We repeat the analysis with the new fitting formula and we find the following constraints on $\alpha_{13}$: $\alpha_{13} = 0.7^{+2.7}_{-2.6}$ from $f_{221}$ and $\alpha_{13} = -2.5^{+3.6}_{-3.1}$ from $\tau_{221}$, both at 90\% CL. To combine these two constraints, we use a kernel density distribution. We find the combined constraint $\alpha_{13} = -0.5^{+2.1}_{-4.0}$ at 90\% CL (orange area in Fig.~\ref{f-fit}).

\section{Concluding remarks}\label{s-c}

In this work, we have calculated the QNMs of a parametric black hole metric in a theory-agnostic approach. We employed the Johannsen metric and we restricted our analysis to the leading order deformation parameters: $\alpha_{13}$, $\alpha_{22}$, $\alpha_{52}$, and $\epsilon_{3}$. We assumed the validity of the Einstein Equations and we consider the approximation of low values of the black hole spin parameter and of the deformation parameters. With such an approximation, we can separate axial and polar modes. Within our approximation, the deformation parameter $\alpha_{22}$ does not appear in the effective potential for axial and polar perturbations.

For the deformation parameter $\alpha_{13}$, we have constructed a fitting formula to describe the frequency and the damping time of the QNMs. We used our fitting formula to analyze the data of GW170104, as the spin of the final BH was found to be low and we see the QNM $\{2,2,1\}$ in the data \cite{LIGOScientific:2020tif}. Our estimate of $\alpha_{13}$ is $-0.5^{+2.1}_{-4.0}$ (90\% CL), which is consistent with no deviations from the Kerr background and is in agreement with the results based on the inspiral phase of the same event reported in Ref.~\cite{Shashank:2021giy}. With the ongoing runs of LVK and the advent of more sensitive future instruments~\cite{LISA:2022yao, Li:2024rnk, Hu:2017mde, Punturo:2010zz, Evans:2021gyd}, we can expect to have new and higher-quality data of events with low final spins for which the method presented in this work can be suitable to test GR in an agnostic way.

The construction of a full waveform model would be certainly possible, but is out of scope of this work. Moreover, here we have only derived a fitting formula in the limit of low spins. However, in reality, the spin of BH binaries in ring-down process are generally larger than $0.3$. Therefore, calculation of high spin value can be important in actual data-fittings. There is a spectral decomposition method~\cite{Bl_zquez_Salcedo_2016,Chung_2024} for high spin calculation which we plan to explore in a subsequent work.

Recently, some authors have proposed Teukolsky-like equations for beyond-GR/beyond-Kerr black holes, which can be used to calculate QNMs of rotating black holes without requiring the slow rotation approximation~\cite{Hussain:2022ins,Li:2022pcy,Cano:2023tmv}. This approach relies on complicated calculations and often requires an expansion in the spin parameter, because rotating black hole solutions of specific gravity theories are typically known perturbatively in the black hole spin parameter. Theory-agnostic tests are easier and more efficient for a null test than a theory-specific approach, which can be applied only to test a specific model and requires higher precision and higher-order calculations.

\vspace{0.5cm}

\begin{acknowledgments}
We would like to thank Gregorio Carullo for answering our queries related to {\tt pyRing}.
This work was supported by the National Natural Science Foundation of China (NSFC), Grant No.~12250610185 and 12261131497, and the Natural Science Foundation of Shanghai, Grant No.~22ZR1403400.
We thank the LVK Collaboration for making their data publicly available.
\end{acknowledgments}

\begin{appendix}  

\section{Effective potential for polar perturbations}\label{A}

In the case of polar perturbations ($\Psi = \tilde{H}^{lm}_0$, $\tilde{H}^{lm}_1$, $\tilde{H}^{lm}_2$, , $\tilde{K}^{lm}$), the effective potential $V$ in Eq.~(\ref{eq-Master}) is 
\begin{widetext}
{\small
\begin{eqnarray}
    &V^{\rm polar}&=\frac{1}{r^4 ((l^2+l-2t) r+6 M)^2}\Bigg(72 (l^2+l-3) M^3 r-(l^2+l-2)^2 r^4 (l^2+l-r^2 \omega ^2) \nonumber \\
    &+&12 M^2 r^2 \left(l^4+2 l^3-6 l^2-7 l+3 r^2 \omega ^2+10\right)+2 \left(l^2+l-2\right) M r^3 \left(l^4+2 l^3-4 l^2-5 l+6 r^2 \omega ^2+6\right)+144 M^4\Bigg) \nonumber \\
    &+&\frac{4 m M a}{l (l+1) r^8 \omega  ((l^2+l-2) r+6 M)^4}\Bigg( (12 (l^2+l-2) M^3 r^4 (l (l+1) (l (l+1) (10 l (l+1)-73)-16 r^2 \omega ^2+240) \nonumber \\
    &-&2 \left(92 r^2 \omega ^2+179\right))+2 \left(l^2+l-2\right)^2 M^2 r^5 \left(l (l+1) \left(l (l+1) (7 l (l+1)-68)-60 r^2 \omega ^2+44\right)-288 r^2 \omega ^2+176\right) \nonumber \\
    &+&\left(l^2+l-2\right)^2 M r^6 \left(-(l-1) (l+2) \left(l^2+l+2\right) (13 l (l+1)-50)-4 (l (l+1) (5 l (l+1)+6)-38) r^2 \omega ^2\right) \nonumber \\
    &+&\left(l^2+l-2\right)^3 r^7 \left(-l^2 (l+1)^2 \left(r^2 \omega ^2-3\right)+8 r^2 \omega ^2-12\right)+1728 (18 l (l+1)-73) M^6 r+288 M^5 r^2 (l (l+1) (21 l (l+1)-202) \nonumber \\
    &+&6 r^2 \omega ^2+401)+24 M^4 r^3 \left(l (l+1) \left(l (l+1) (29 l (l+1)-384)+6 \left(3 r^2 \omega ^2+277\right)\right)-4 \left(45 r^2 \omega ^2+532\right)\right)+55296 M^7)\Bigg) \nonumber \\
    &+&\frac{\alpha_{13} M^3}{r^7 \left(\left(l^2+l-2\right) r+6 M\right)^3}\Bigg( (-12 (l^2+l-2) M r^4 (l (l+1) (8 l (l+1)-4 r^2 \omega ^2+47)+14 r^2 \omega ^2-195) \nonumber \\
    &+&(l^2+l-2)^2 r^5 \left(2 \left(l^2+l-5\right) r^2 \omega ^2+21 l (l+1)+120\right)-72 (140 l (l+1)-661) M^4 r+36 M^3 r^2 (4 \left(6 r^2 \omega ^2-293\right) \nonumber \\
    &-&l (l+1) (25 l (l+1)-486))+6 M^2 r^3 \left(l (l+1) \left(l (l+1) (18 l (l+1)+175)+60 r^2 \omega ^2-1930\right)-156 r^2 \omega ^2+3232\right)-22032 M^5)\Bigg) \nonumber \\
    &+&\frac{\alpha_{52} M^2}{2 r^6 \left(\left(l^2+l-2\right) r+6 M\right)^4}\Bigg( (-4 \left(l^2+l-2\right) M^2 r^4 \left(l (l+1) \left(l (l+1) (7 l (l+1)+233)+8 \left(3 r^2 \omega ^2-262\right)\right)+132 r^2 \omega ^2+3384\right) \nonumber \\
    &-&4 \left(l^2+l-2\right)^2 M r^5 \left(l (l+1) \left((l-1) l (l+1) (l+2)+6 r^2 \omega ^2-91\right)+30 r^2 \omega ^2+236\right)+\left(l^2+l-2\right)^3 r^6 (\left(l^2+l-2\right) (2 l (l+1) \nonumber \\
    &+&11)-2 \left(l^2+l+4\right) r^2 \omega ^2)+288 (205 l (l+1)-689) M^5 r+144 M^4 r^2 \left(31 l (l+1) (3 l (l+1)-25)-6 r^2 \omega ^2+1319\right) \nonumber \\
    &+&8 M^3 r^3 \left(l (l+1) \left(l (l+1) (101 l (l+1)-2415)-36 r^2 \omega ^2+10410\right)-4 \left(9 r^2 \omega ^2+3019\right)\right)+86400 M^6)\Bigg) \nonumber \\
    &-&\frac{3  \epsilon_3 M^3 (2 M-r)}{r^7 \left(\left(l^2+l-2\right) r+6 M\right)^4}\Bigg(12 \left(l^2+l-2\right) M^2 r^3 \left(l (l+1) (37 l (l+1)-317)+60 r^2 \omega ^2+552\right) \nonumber \\
    &+&\left(l^2+l-2\right)^2 M r^4 \left(l (l+1) (19 l (l+1)-254)+6 \left(22 r^2 \omega ^2+93\right)\right)-8 \left(l^2+l-2\right)^3 r^5 \left(l^2+l-r^2 \omega ^2-2\right)\nonumber \\
    &+&144 (131 l (l+1)-358) M^4 r+12 M^3 r^2 \left(7 l (l+1) (53 l (l+1)-320)+4 \left(27 r^2 \omega ^2+776\right)\right)+28512 M^5\Bigg)
\end{eqnarray}
}
\end{widetext}
As in the case of the effective potential for axial perturbations, within our approximation we do not have the deformation parameter $\alpha_{22}$.

\begin{figure}
    \centering
    \includegraphics[width=0.9\textwidth]{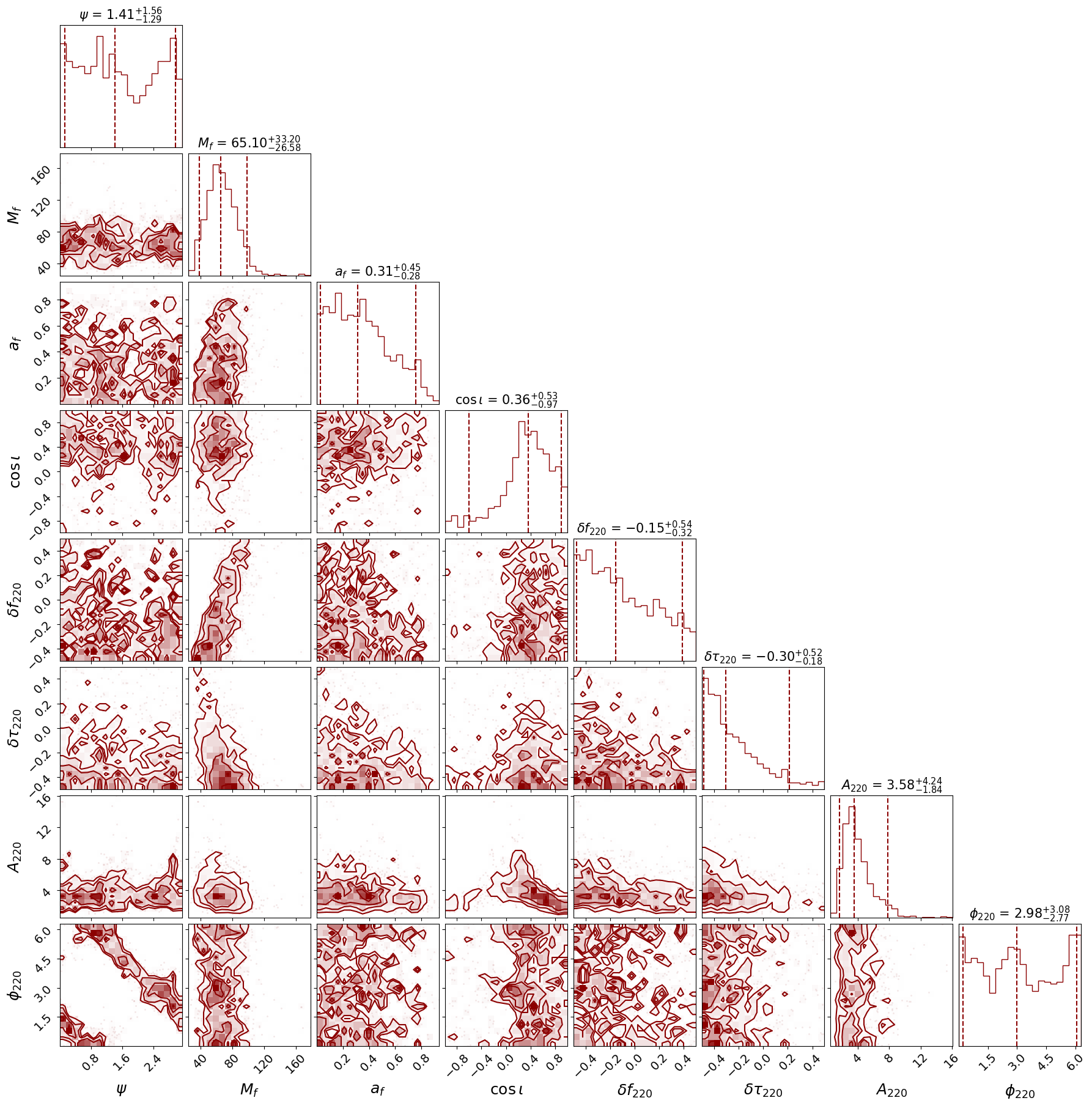}
    \caption{Posteriors for {\tt pyRing} analysis of GW170104.}\label{f-posterior}
\end{figure}

\begin{widetext}
\section{{\tt pyRing} analysis initial settings}
We use {\tt pyRing} to analyze the ringdown data of event GW170104. We use default prior values of Parspec formalism in {\tt pyRing}. The ringdown data used is publicly available at https://dcc.ligo.org/LIGO-P2000438/public.
In this appendix, we present the settings for the {\tt pyRing} setup and the final posteriors in Fig.~\ref{f-posterior}.
\begin{longtable}{|l|l|}
\hline
\textbf{Parameter} & \textbf{Value} \\ \hline
\endfirsthead
\hline
\textbf{Parameter} & \textbf{Value} \\ \hline
\endhead
\hline
\endfoot
\hline
\endlastfoot
\textbf{Input Settings} & \\ \hline
run-type & full \\ \hline
pesummary & 0 \\ \hline
screen-output & 1 \\ \hline
output & gw170104\_ParSpec \\ \hline
data-H1 & gwdata/H-H1\_GWOSC\_4KHZ\_R1-1167559921-32.txt \\ \hline
data-L1 & gwdata/L-L1\_GWOSC\_4KHZ\_R1-1167559921-32.txt \\ \hline
trigtime & 1167559936.59363 \\ \hline
detectors & H1, L1 \\ \hline
sky-frame & equatorial \\ \hline
kerr-modes & [(2,2,2,0)] \\ \hline
reference-amplitude & 1E-21 \\ \hline
amp-non-prec-sym & 1 \\ \hline
domega-tgr-modes & [(2,2,0)] \\ \hline
dtau-tgr-modes & [(2,2,0)] \\ \hline
ParSpec & 1 \\ \hline
ParSpec\_Dmax\_TGR & 0 \\ \hline
ParSpec\_Dmax\_charge & 0 \\ \hline
\textbf{Sampler Settings} & \\ \hline
nlive & 256 \\ \hline
maxmcmc & 256 \\ \hline
seed & 1234 \\ \hline
\textbf{Priors} & \\ \hline
fix-t & 0.0027673 \\ \hline
fix-ra & 2.008144 \\ \hline
fix-dec & 0.144163 \\ \hline
fix-phi & 0.0 \\ \hline
\end{longtable}
\end{widetext}

\section{Deformed ring-down waveform}\label{B}

\begin{figure*}[t]
    \centering
    \includegraphics[width=0.98\textwidth]{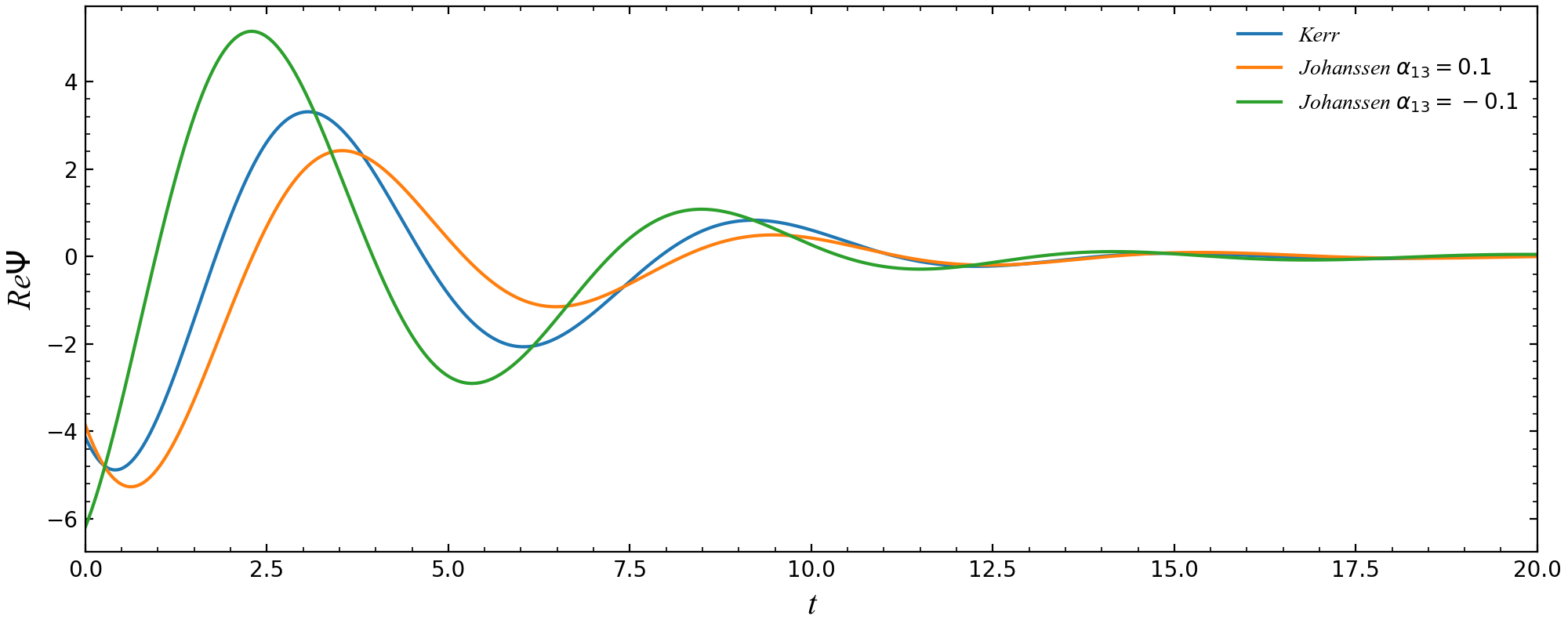}
    \caption{Comparison among radial ring-down waveforms Kerr and Johannsen spacetimes.}\label{f-wf}
    \end{figure*}

In Fig.~\ref{f-wf}, we show the radial ring-down waveform of Kerr and Johannsen spacetime with $a=0.35$ and $\alpha_{13}=-0.1$, 0, and 0.1. The waveform combines the modes $(l,|m|)=(2,1)$, (2,2), (3,3), (4,4), and (5,5). For the amplitude of every mode, we adopt the analytical expression of the amplitude in Ref.~\cite{PhysRevD.90.124032}. We set the masses of the two progenitor BHs in the binary system to $m_{1}=M$ and $m_{2}=2\,M$, and we plot the oscillation of the waveform for a point at $r = 10 \, M$. In waveform construction, we assume that the analytical expression which relatee the amplitude and certain QNM frequencies are the same for the Kerr and Johannsen metrics. The calculation of QNM frequencies are all based on the root finder in Sec.~\ref{s-qnm0}

The ring-down waveform model follows following analytical formula
\begin{eqnarray}
    h_{+}-ih_{\times }=\frac{M}{r}\sum_{lmn}\mathcal{A}_{lmn} S_{lmn}(\theta,\phi)e^{i \omega_{lmn}t}e^{-t/\tau_{lmn}}
\end{eqnarray}
For the radial function in equatorial plane ($(\theta,\phi)=(0,0)$), the relative amplitude of the real part waveform follows
\begin{eqnarray}
    h_{+}-ih_{\times }=h\sim \sum_{lmn}\mathcal{A}_{lmn} e^{i \omega_{lmn}t}e^{-t/\tau_{lmn}} \, .
\end{eqnarray}
The amplitude $\mathcal{A}_{lmn}$ is discussed in Ref.~\cite{PhysRevD.90.124032}.
\begin{eqnarray}
    {A}_{lmn}=\eta \omega^{2}_{lmn}\sum_{n}a_{n}e^{i\alpha_{n}}\eta^{n}
\end{eqnarray}
where
\begin{eqnarray}
    \eta=\frac{m_{1}m_{2}}{(m_{1}+m_{2})^{2}} \, ,
\end{eqnarray}
and $a_{n}$ and $\alpha_{n}$ are fitting amplitude and phase factor respectfully for complex wave function value.

\section{Quasi normal frequencies for different $(l,m)$}
\label{C}
We present different QNM frequencies for different $(l,m)$. The complex frequency value is presented with the form (Re($\omega$),-Im($\omega$)).

\begin{table*}[h]
    \caption{QNM frequencies for $l=2$, $m=-2,-1,0,1,2$
    }
    \begin{tabular}{ccccccc}
        \hline
        \hline
       $a/M$&$\alpha_{13}$&$m=-2$&$m=-1$&$m=0$&$m=1$&$m=2$ \\ \hline

    \multirow{7}{*}{0} & -0.3&  0.3795, 0.0929  \qquad& 0.3795, 0.0929  \qquad& 0.3795, 0.0929 \qquad& 0.3795, 0.0929 \qquad& 0.3796, 0.0929 \\
                       & -0.2&  0.3775, 0.0915  \qquad& 0.3775, 0.0915  \qquad& 0.3775, 0.0915 \qquad& 0.3775, 0.0915 \qquad& 0.3776, 0.0915 \\
                       & -0.1&  0.3756, 0.0902  \qquad& 0.3756, 0.0902  \qquad& 0.3756, 0.0902 \qquad& 0.3756, 0.0902 \qquad& 0.3757, 0.0902 \\
                        \qquad& 0.0&  0.3736, 0.0889  \qquad& 0.3736, 0.0889  \qquad& 0.3736, 0.0889 \qquad& 0.3736, 0.0889 \qquad& 0.3737, 0.0889 \\
                        \qquad& 0.1&  0.3717, 0.0876  \qquad& 0.3717, 0.0876  \qquad& 0.3717, 0.0876 \qquad& 0.3717, 0.0876 \qquad& 0.3718, 0.0876 \\
                        \qquad& 0.2&  0.3798, 0.0864  \qquad& 0.3698, 0.0864  \qquad& 0.3698, 0.0864 \qquad& 0.3698, 0.0864 \qquad& 0.3699, 0.0864 \\
                        \qquad& 0.3&  0.3679, 0.0851  \qquad& 0.3679, 0.0851  \qquad& 0.3679, 0.0851 \qquad& 0.3679, 0.0851 \qquad& 0.3680, 0.0851 \\ \hline
    \multirow{7}{*}{0.1}& -0.3&  0.3675, 0.0926  \qquad& 0.3736, 0.0927  \qquad& 0.3795, 0.0929 \qquad& 0.3863, 0.0929 \qquad& 0.3930, 0.0931 \\
                       & -0.2&  0.3656, 0.0915  \qquad& 0.3716, 0.0914  \qquad& 0.3775, 0.0915 \qquad& 0.3843, 0.0915 \qquad& 0.3910, 0.0916 \\
                       & -0.1&  0.3636, 0.0903  \qquad& 0.3697, 0.0902  \qquad& 0.3756, 0.0902 \qquad& 0.3823, 0.0901 \qquad& 0.3890, 0.0901 \\
                        \qquad& 0.0&  0.3618, 0.0891  \qquad& 0.3678, 0.0890  \qquad& 0.3736, 0.0889 \qquad& 0.3804, 0.0887 \qquad& 0.3870, 0.0887 \\
                        \qquad& 0.1&  0.3599, 0.0888  \qquad& 0.3659, 0.0878  \qquad& 0.3717, 0.0876 \qquad& 0.3784, 0.0874 \qquad& 0.3850, 0.0873 \\
                        \qquad& 0.2&  0.3580, 0.0869  \qquad& 0.3640, 0.0866  \qquad& 0.3698, 0.0864 \qquad& 0.3765, 0.0861 \qquad& 0.3831, 0.0859 \\
                        \qquad& 0.3&  0.3562, 0.0858  \qquad& 0.3621, 0.0854  \qquad& 0.3679, 0.0851 \qquad& 0.3746, 0.0848 \qquad& 0.3812, 0.0846 \\ \hline
    \multirow{7}{*}{0.2} & -0.3&  0.3560, 0.0922  \qquad& 0.3684, 0.0924  \qquad& 0.3795, 0.0929 \qquad& 0.3912, 0.0928 \qquad& 0.4078, 0.0929 \\
                       & -0.2&  0.3547, 0.0912  \qquad& 0.3665, 0.0913  \qquad& 0.3775, 0.0915 \qquad& 0.3922, 0.0913 \qquad& 0.4059, 0.0913 \\
                       & -0.1&  0.3529, 0.0901  \qquad& 0.3646, 0.0901  \qquad& 0.3756, 0.0902 \qquad& 0.3902, 0.0898 \qquad& 0.4040, 0.0898 \\
                        \qquad& 0.0&  0.3511, 0.0891  \qquad& 0.3627, 0.0889  \qquad& 0.3736, 0.0889 \qquad& 0.3882, 0.0884 \qquad& 0.4021, 0.0883 \\
                        \qquad& 0.1&  0.3493, 0.0881  \qquad& 0.3608, 0.0878  \qquad& 0.3717, 0.0876 \qquad& 0.3863, 0.0870 \qquad& 0.4002, 0.0868 \\
                        \qquad& 0.2&  0.3475, 0.0872  \qquad& 0.3590, 0.0867  \qquad& 0.3698, 0.0864 \qquad& 0.3843, 0.0856 \qquad& 0.3983, 0.0854 \\
                        \qquad& 0.3&  0.3457, 0.0862  \qquad& 0.3571, 0.0856  \qquad& 0.3679, 0.0851 \qquad& 0.3824, 0.0843 \qquad& 0.3964, 0.0840 \\ \hline
    \multirow{7}{*}{0.3}& -0.3&  0.3465, 0.0918  \qquad& 0.3639, 0.0920  \qquad& 0.3795, 0.0929 \qquad& 0.4032, 0.0925 \qquad& 0.4236, 0.0918 \\
                       & -0.2&  0.3448, 0.0909  \qquad& 0.3620, 0.0909  \qquad& 0.3775, 0.0915 \qquad& 0.4012, 0.0909 \qquad& 0.4224, 0.0904 \\
                       & -0.1&  0.3430, 0.0090  \qquad& 0.3602, 0.0898  \qquad& 0.3756, 0.0902 \qquad& 0.3992, 0.0894 \qquad& 0.4210, 0.0890 \\
                        \qquad& 0.0&  0.3413, 0.0892  \qquad& 0.3583, 0.0887  \qquad& 0.3736, 0.0889 \qquad& 0.3973, 0.0879 \qquad& 0.4195, 0.0877 \\
                        \qquad& 0.1&  0.3395, 0.0883  \qquad& 0.3565, 0.0877  \qquad& 0.3717, 0.0876 \qquad& 0.3953, 0.0865 \qquad& 0.4179, 0.0865 \\
                        \qquad& 0.2&  0.3378, 0.0875  \qquad& 0.3546, 0.0866  \qquad& 0.3698, 0.0864 \qquad& 0.3934, 0.0851 \qquad& 0.4162, 0.0852 \\
                        \qquad& 0.3&  0.3361, 0.0866  \qquad& 0.3528, 0.0856  \qquad& 0.3679, 0.0851 \qquad& 0.3815, 0.0837 \qquad& 0.4144, 0.0840 \\ \hline
     \hline
    \end{tabular}
    \end{table*}

\begin{table*}
    \caption{Fitting coefficients for the real parts of $l=2$ QNM frequencies.}
    \begin{tabular}{cccc}
        \hline
        \hline
       $m$&$P_{0}$&$P_{1}$& error [\%] \\ \hline

    -2 &-0.0195 \qquad &0.0071 \qquad &  0.19 \\ \hline
    -1 &-0.0194 \qquad&0.0028 \qquad&    0.19 \\ \hline
    0 &-0.0193 \qquad&0.0000 \qquad&   0.20 \\ \hline
    1 &-0.0194 \qquad&-0.0007 \qquad&  0.20 \\ \hline
    2 &-0.0197 \qquad&0.0025 \qquad&  0.18\\ \hline
    \end{tabular}
    \end{table*}

\begin{table*}
        \caption{Fitting coefficients for the imaginary parts of $l=2$ QNM frequencies.}
        \begin{tabular}{cccc}
            \hline
            \hline
           $m$&$Q_{0}$&$Q_{1}$& error [\%] \\ \hline
    
    -2 &0.0129 \qquad&-0.0144 \qquad&    0.11 \\ \hline
    -1 &0.0129 \qquad&-0.0075 \qquad&    0.12 \\ \hline
    0 &0.0129 \qquad&0.0000 \qquad&   0.13 \\ \hline
    1 &0.0129 \qquad&0.0057 \qquad&  0.14 \\ \hline
    2 &0.0130 \qquad&-0.0095 \qquad&  0.14\\ \hline
        \end{tabular}
        \end{table*}

\begin{table*}
    \caption{QNM frequencies for $l=3$, $m=-3,-2,-1,0,1,2,3$
    }
    \begin{tabular}{ccccccccc}
        \hline
        \hline
       $a/M$&$\alpha_{13}$&$m=-3$&$m=-2$&$m=-1$&$m=0$&$m=1$&$m=2$&$m=3$ \\ \hline
    \multirow{7}{*}{0} & -0.3&  0.6066, 0.0947  \qquad& 0.6066, 0.0947  \qquad& 0.6066, 0.0947 \qquad& 0.6066, 0.0947 \qquad& 0.6066, 0.0947 \qquad& 0.6066, 0.0947 \qquad& 0.6066, 0.0947 \\
                       & -0.2&  0.6042, 0.0940  \qquad& 0.6042, 0.0940  \qquad& 0.6042, 0.0940 \qquad& 0.6042, 0.0940 \qquad& 0.6042, 0.0940 \qquad& 0.6042, 0.0940 \qquad& 0.6042, 0.0940 \\
                       & -0.1&  0.6018, 0.0933  \qquad& 0.6018, 0.0933  \qquad& 0.6018, 0.0933 \qquad& 0.6018, 0.0933 \qquad& 0.6018, 0.0933 \qquad& 0.6018, 0.0933 \qquad& 0.6018, 0.0933 \\
                        \qquad& 0.0&  0.5994, 0.0927  \qquad& 0.5994, 0.0927  \qquad& 0.5994, 0.0927 \qquad& 0.5994, 0.0927 \qquad& 0.5994, 0.0927 \qquad& 0.5994, 0.0927 \qquad& 0.5994, 0.0927 \\
                        \qquad& 0.1&  0.5970, 0.0920  \qquad& 0.5970, 0.0920  \qquad& 0.5970, 0.0920 \qquad& 0.5970, 0.0920 \qquad& 0.5970, 0.0920 \qquad& 0.5970, 0.0920 \qquad& 0.5970, 0.0920 \\
                        \qquad& 0.2&  0.5947, 0.0914  \qquad& 0.5947, 0.0914  \qquad& 0.5947, 0.0914 \qquad& 0.5947, 0.0914 \qquad& 0.5947, 0.0914 \qquad& 0.5947, 0.0914 \qquad& 0.5947, 0.0914 \\
                        \qquad& 0.3&  0.5924, 0.0907  \qquad& 0.5924, 0.0907  \qquad& 0.5924, 0.0907 \qquad& 0.5924, 0.0907 \qquad& 0.5924, 0.0907 \qquad& 0.5924, 0.0907 \qquad& 0.5924, 0.0907\\ \hline
    \multirow{7}{*}{0.1}& -0.3&  0.5867, 0.0947  \qquad& 0.5933, 0.0946  \qquad& 0.6002, 0.0947 \qquad& 0.6066, 0.0947 \qquad& 0.6142, 0.0946 \qquad& 0.6216, 0.0945 \qquad& 0.6291, 0.0945 \\
                       & -0.2&  0.5845, 0.0941  \qquad& 0.5910, 0.0940  \qquad& 0.5978, 0.0940 \qquad& 0.6042, 0.0940 \qquad& 0.6117, 0.0939 \qquad& 0.6189, 0.0938 \qquad& 0.6263, 0.0938 \\
                       & -0.1&  0.5824, 0.0935  \qquad& 0.5888, 0.0933  \qquad& 0.5955, 0.0933 \qquad& 0.6018, 0.0933 \qquad& 0.6092, 0.0932 \qquad& 0.6163, 0.0932 \qquad& 0.6235, 0.0931 \\
                        \qquad& 0.0&  0.5802, 0.0928  \qquad& 0.5866, 0.0927  \qquad& 0.5932, 0.0927 \qquad& 0.5994, 0.0927 \qquad& 0.6067, 0.0926 \qquad& 0.6137, 0.0925  \qquad& 0.6208, 0.0924 \\
                        \qquad& 0.1&  0.5781, 0.0922  \qquad& 0.5844, 0.0921  \qquad& 0.5909, 0.0921 \qquad& 0.5970, 0.0920 \qquad& 0.6042, 0.0919 \qquad& 0.6112, 0.0918  \qquad& 0.6181, 0.0918 \\
                        \qquad& 0.2&  0.5761, 0.0916  \qquad& 0.5823, 0.0914  \qquad& 0.5887, 0.0914 \qquad& 0.5947, 0.0914 \qquad& 0.6018, 0.0913 \qquad& 0.6086, 0.0912  \qquad& 0.6155, 0.0911\\
                        \qquad& 0.3&  0.5740, 0.0910  \qquad& 0.5801, 0.0908  \qquad& 0.5865, 0.0908 \qquad& 0.5924, 0.0907 \qquad& 0.5995, 0.0906 \qquad& 0.6062, 0.0905  \qquad& 0.6129, 0.0905 \\ \hline
    \multirow{7}{*}{0.2} & -0.3&  0.5685, 0.0946  \qquad& 0.5814, 0.0945  \qquad& 0.5947, 0.0945 \qquad& 0.6066, 0.0947 \qquad& 0.6231, 0.0943 \qquad& 0.6393, 0.0941 \qquad& 0.6544, 0.0938 \\
                       & -0.2&  0.5666, 0.0940  \qquad& 0.5792, 0.0939  \qquad& 0.5924, 0.0939 \qquad& 0.6042, 0.0940 \qquad& 0.6204, 0.0936 \qquad& 0.6354, 0.0934 \qquad& 0.6512, 0.0932 \\
                       & -0.1&  0.5646, 0.0934  \qquad& 0.5771, 0.0903  \qquad& 0.5902, 0.0932 \qquad& 0.6018, 0.0933 \qquad& 0.6178, 0.0930 \qquad& 0.6325, 0.0928 \qquad& 0.6480, 0.0926 \\
                        \qquad& 0.0&  0.5627, 0.0928  \qquad& 0.5751, 0.0927  \qquad& 0.5880, 0.0926 \qquad& 0.5994, 0.0927 \qquad& 0.6152, 0.0923 \qquad& 0.6297, 0.0921  \qquad& 0.6448, 0.0920 \\
                        \qquad& 0.1&  0.5608, 0.0923  \qquad& 0.5731, 0.0921  \qquad& 0.5858, 0.0920 \qquad& 0.5970, 0.0920 \qquad& 0.6127, 0.0916 \qquad& 0.6269, 0.0914  \qquad& 0.6417, 0.0914 \\
                        \qquad& 0.2&  0.5590, 0.0917  \qquad& 0.5711, 0.0915  \qquad& 0.5837, 0.0913 \qquad& 0.5947, 0.0914 \qquad& 0.6101, 0.0910 \qquad& 0.6241, 0.0908  \qquad& 0.6387, 0.0908 \\
                        \qquad& 0.3&  0.5571, 0.0912  \qquad& 0.5691, 0.0909  \qquad& 0.5815, 0.0907 \qquad& 0.5924, 0.0907 \qquad& 0.6077, 0.0903 \qquad& 0.6214, 0.0902  \qquad& 0.6357, 0.0902 \\ \hline
    \multirow{7}{*}{0.3} & -0.3&  0.5522, 0.0943  \qquad& 0.5605, 0.0944  \qquad& 0.5902, 0.0942 \qquad& 0.6066, 0.0947 \qquad& 0.6335, 0.0939 \qquad& 0.6575, 0.0933 \qquad& 0.6845, 0.0922 \\
                       & -0.2&  0.5504, 0.0938  \qquad& 0.5686, 0.0938  \qquad& 0.5880, 0.0936 \qquad& 0.6042, 0.0940 \qquad& 0.6307, 0.0932 \qquad& 0.6543, 0.0927 \qquad& 0.6805, 0.0919 \\
                       & -0.1&  0.5487, 0.0933  \qquad& 0.5666, 0.0932  \qquad& 0.5859, 0.0930 \qquad& 0.6018, 0.0933 \qquad& 0.6279, 0.0925 \qquad& 0.6511, 0.0921 \qquad& 0.6766, 0.0916 \\
                        \qquad& 0.0&  0.5469, 0.0927  \qquad& 0.5647, 0.0926  \qquad& 0.5837, 0.0923 \qquad& 0.5994, 0.0927 \qquad& 0.6252, 0.0918 \qquad& 0.6479, 0.0915  \qquad& 0.6721, 0.0913\\
                        \qquad& 0.1&  0.5452, 0.0922  \qquad& 0.5628, 0.0921  \qquad& 0.5816, 0.0917 \qquad& 0.5970, 0.0920 \qquad& 0.6225, 0.0912 \qquad& 0.6448, 0.0909  \qquad& 0.6692, 0.0909 \\
                        \qquad& 0.2&  0.5435, 0.0917  \qquad& 0.5610, 0.0915  \qquad& 0.5796, 0.0911 \qquad& 0.5947, 0.0914 \qquad& 0.6199, 0.0905 \qquad& 0.6418, 0.0903  \qquad& 0.6656, 0.0905 \\
                        \qquad& 0.3&  0.5418, 0.0912  \qquad& 0.5591, 0.0910  \qquad& 0.5775, 0.0905 \qquad& 0.5924, 0.0907 \qquad& 0.6173, 0.0899 \qquad& 0.6388, 0.0897  \qquad& 0.6620, 0.0901 \\ \hline
     \hline
    \end{tabular}
    \end{table*}

\begin{table*}
    \caption{Fitting coefficients for the real parts of $l=3$ QNM frequencies.
    }
    \begin{tabular}{cccc}
        \hline
        \hline
        $m$&$P_{0}$&$P_{1}$ & error [\%] \\ \hline

    -3 &  -0.0023 & 0.0213  \qquad& 0.21 \\ \hline
    -2 &  -0.0024 & 0.0156  \qquad& 0.22 \\ \hline
    -1 &  -0.0024 & 0.0086  \qquad& 0.23 \\ \hline
    0 &  -0.0024 & -0.0000  \qquad& 0.24 \\ \hline
    1 &  -0.0024 & -0.0109  \qquad& 0.26 \\ \hline
    2 &  -0.0023 & -0.0250  \qquad& 0.28 \\ \hline
    3 &  -0.0023 & -0.0453  \qquad& 0.30 \\ \hline
    \end{tabular}
    \end{table*}
        
\begin{table*}
    \caption{Fitting coefficients for the imaginary parts of $l=3$ QNM frequencies.
    }
    \begin{tabular}{cccc}
        \hline
        \hline
        $m$&$Q_{0}$&$Q_{1}$& error [\%] \\ \hline

    -3 &  0.0066 & -0.0047  \qquad& 0.06 \\ \hline
    -2 &  0.0066 & -0.0042  \qquad& 0.07 \\ \hline
    -1 &  0.0066 & -0.0013  \qquad& 0.06 \\ \hline
    0 &  0.0066 & 0.0000  \qquad& 0.07 \\ \hline
    1 &  0.0066 & 0.0003  \qquad& 0.07 \\ \hline
    2 &  0.0067 & -0.0019  \qquad& 0.07 \\ \hline
    3 &  0.0067 & -0.0030  \qquad& 0.06 \\ \hline
    \end{tabular}
    \end{table*}
\end{appendix}

\bibliography{references}

\end{document}